\documentclass[10pt]{iopart}

\usepackage[USenglish]{babel} 
\usepackage[utf8x]{inputenc}
\usepackage{lmodern} 
\usepackage{geometry}
\usepackage{float}
\usepackage[pdftex]{graphicx}
\usepackage[bf,small,format=plain]{caption}  

\newcommand{\text}[1]{{\mbox{\scriptsize #1}}}
\newcommand{\eqref}[1]{(\ref{#1})}
\newcommand{\TE}{TE$_{01}${ }}
\newcommand{\TM}{TM$_{01}${ }}
\newcommand{\HE}{HE$_{11}${ }}
\begin{document}

\title[Optimized Single-Crystal Diamond Scanning Probes for High Sensitivity Magnetometry]{Optimized Single-Crystal Diamond Scanning Probes for High Sensitivity Magnetometry}

\author{P Fuchs$^1$, M Challier$^1$ and E Neu$^1$}
\address{$^1$ Faculty of Natural Sciences and Technology, Physics, Saarland University, 66123 Saarbrücken, Germany}
\begin{abstract}
The negatively-charged nitrogen-vacancy center (NV) in diamond forms a versatile system for quantum sensing applications. Combining the advantageous properties of this atomic-sized defect with scanning probe techniques such as atomic force microscopy (AFM) enables nanoscale imaging of e.g.\ magnetic fields. To form a scanning probe device, we place single NVs shallowly (i.e.\ $< 20$ nm) below the top facet of a diamond nanopillar, which is located on a thin diamond platform of typically below $1~\mu$m thickness. This device can be attached to an AFM head, forming an excellent scanning probe tip. Furthermore, it simultaneously influences the collectible photoluminescence (PL) rate of the NV located inside. Especially sensing protocols using continuous optically-detected magnetic resonance (ODMR) benefit from an enhanced collectible PL rate, improving the achievable sensitivity. 
This work presents a comprehensive set of simulations to quantify the influence of the device geometry on the collectible PL rate for individual NVs. Besides geometric parameters (e.g.\ pillar length, diameter and platform thickness), we also focus on fabrication uncertainties such as the exact position of the NV or the taper geometry of the pillar introduced by imperfect etching. As a last step, we use these individual results to optimize our current device geometry, yielding a realistic gain in collectible PL rate by a factor of 13 compared to bulk diamond and 1.8 compared to our unoptimized devices.
\end{abstract}
\maketitle

\section{Introduction}
In the last years, diamond has emerged as a very promising host material for the implementation of single, atomic-sized quantum systems \cite{Atature2018}. A variety of optically-active point defects were found, with many of them exhibiting discrete energy levels suitable for applications in quantum technologies. Especially the negatively-charged nitrogen-vacancy center (NV) in diamond, a defect which consists of a vacancy adjacent to a substitutional nitrogen atom \cite{Davies1976}, has been identified as a favorable system in quantum information \cite{Jelezko2002,Tamarat2008} as well as in quantum sensing \cite{Bernardi2017}. Single NVs in diamond possess a bright photoluminescence (PL) and a remarkable photostability \cite{Gruber1997a}. Additionally, their internal population dynamics allow an optical readout of their electronic spin state, so-called optically-detected magnetic resonance (ODMR)\cite{Rondin2014}.
This enables measuring temperatures \cite{Kucsko2013} and pressures \cite{Doherty2014} as well as electric \cite{Dolde2014} and magnetic \cite{Rondin2014,Maze2008} fields with outstanding sensitivity, even at ambient conditions or in biological systems \cite{Schirhagl2014}. 
However, as long as these NVs are observed with a confocal microscope, individual sensing points (i.e.\ single NVs) have to be separated by more than the optical resolution ($\approx 400$ nm).
To enable truly nanoscale imaging, controlled scanning of the NV in close proximity to the sample via a scanning probe technique such as atomic force microscopy (AFM) is mandatory. An NV placed shallowly ($< 20$ nm) below the top facet of a diamond nanopillar forms an excellent scanning probe for nanoscale sensing and imaging \cite{Maletinsky2012,Kleinlein2016,Appel2016}. 
In such a design, the nanopillar serves not only as a tip, enabling a controlled approach of the NV to the sample, but also as a nanophotonic device, guiding the PL of the NV efficiently towards the collection optics. Especially for sensing protocols based on continuous ODMR, it is crucial to maximize collectible PL rate to achieve a high sensitivity \cite{Rondin2014}.
Previous research has mainly focused on demonstrating optimized geometries for nanopillars on bulk diamond to significantly improve the achievable collection efficiency \cite{Neu2014b,Widmann2015,Momenzadeh2015,Marseglia2018}. 
For miniaturized scanning probes (overall size $< 30~\mu$m), however, it can be highly-advantageous to replace the bulk diamond substrate by a thin ($< 1~\mu$m) diamond platform \cite{Appel2016}. First commercial diamond scanning probes follow that design concept of miniaturization \cite{qnami,qzabre}, while for larger scanning probe devices, thicker platforms ($\approx 50~\mu$m) have been used \cite{Zhou2017}. Thin platforms, in general, ease mounting of the probe to an AFM head, but simultaneously alter the photonic properties of the device. 
In this sense, this work gives a comprehensive set of simulations to quantify the influence of several geometric parameters of the scanning probe device on the collectible PL rate. 
The manuscript is structured as follows: Section 2 and 3 shortly introduce NV based magnetometry and our simulation setup. In section 4, we start by generally identifying the influence of different geometrical aspects of our device on the collectible PL rate, namely the position and orientation of the NV inside the pillar, the diameter and length of the pillar as well as the platform thickness. Because our fabricated pillars do not resemble a perfect cylindrical shape, but a truncated cone with trenches forming around the pillar, we also take these two features into account. Finally, we incorporate findings from our nanofabrication to optimize our device geometry for a maximal collectible PL rate and determine the performance of realistic devices.
\section{Magnetometry with NVs in diamond}
\label{sec:magnetometry}
\begin{figure}[h]
\centering
\includegraphics[width=\linewidth]{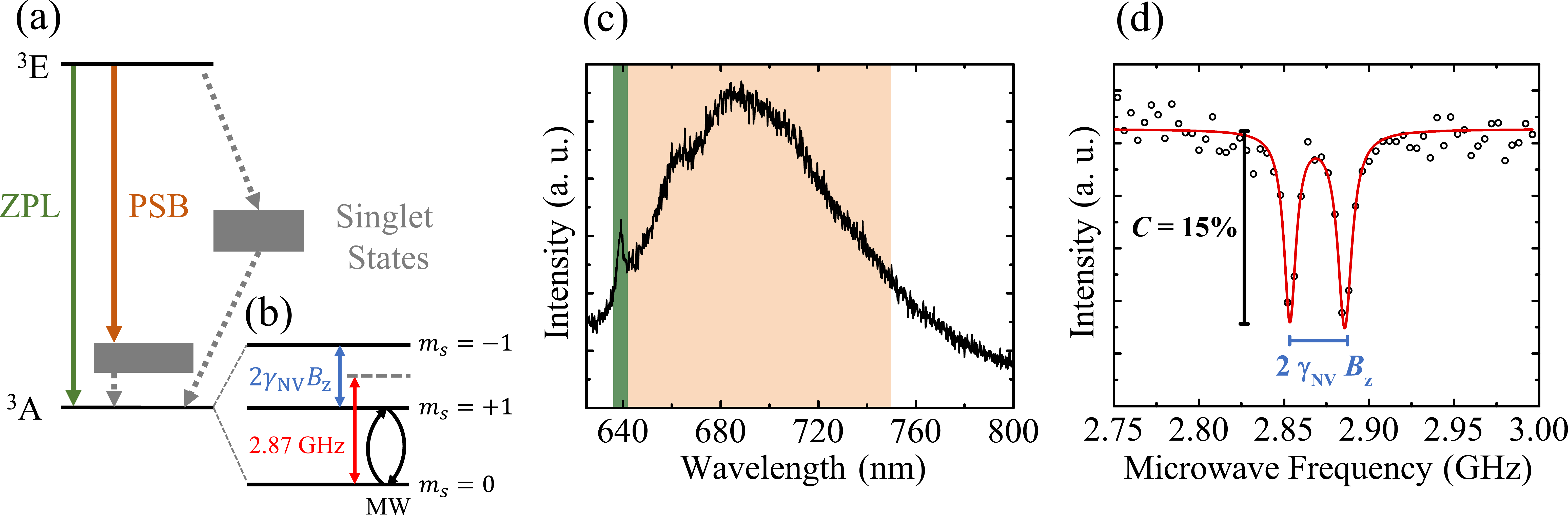}
\caption{The electronic level structure of the NV (a) shows the transitions forming the purely-electronic ZPL as well as the PSB, which originates from the decay of the excited state into the vibrationally-excited ground states. These transitions are all electric dipole transitions, leading to a typical spectrum (c) with a narrow ZPL and a broad PSB. 
Excited state and ground state are both triplet states, whereas the Zeeman splitting of the ground state can be used to detect external magnetic fields (b). A typical curve of such an ODMR measurement (d) shows two dips in PL symmetrically-aligned around the zero field splitting of $2.87$ GHz. The distance of both peaks is proportional to the projection of the external magnetic field on the NV symmetry axis $B_z$.}
\label{fig:NVSpec_ODMR}
\end{figure}

One of the main advantages of the NV is the possibility to perform an optical readout of its electronic spin state, paving the way for many sensing applications. To explain the principle behind ODMR, figure \ref{fig:NVSpec_ODMR} (a) shows a simplified level structure of the NV and figure \ref{fig:NVSpec_ODMR} (c) its PL spectrum at room temperature:
The purely electronic transition between an excited state ($^3E$) and a ground state ($^3A$) leads to the zero-phonon line (ZPL) at 638 nm. Additionally, decays to vibrationally-excited ground states induce a broad, red-shifted phonon sideband (PSB), which spans a bandwidth of about 100 nm and contains around 96 \% of the NV PL \cite{Bernien2012}.

Ground and excited state form a triplet with $m_s = 0$ and $m_s = \pm 1$ substates. In absence of external fields, the splitting between $m_s = 0$ and $m_s = \pm 1$ is $2.87$ GHz for $^3A$; the $m_s = \pm 1$ states are degenerate.

In addition to the direct decay from $^3A$ to $^3E$, a decay channel involving a long-lived singlet state exists. The probability for a decay from $^3E$ via this intersystem crossing is higher for an NV in $m_s = \pm 1$ than for $m_s = 0$, where it is more likely to undergo cycling transitions between $^3A$ and $^3E$. These internal population dynamics allow to polarize the electronic spin of the NV via off-resonant excitation (typically at 532 nm, 1 $\mu$s pulse duration) and also lead to a lower PL rate in the $m_s = \pm 1$ states compared to the $m_s = 0$ state for off-resonant, continuous wave (cw) excitation.
\par
External magnetic fields lift the degeneracy of the $m_s = \pm 1$ levels, as depicted in figure \ref{fig:NVSpec_ODMR} (b). 
Off-resonantly exciting the NV and simultaneously sweeping the frequency of a microwave over the transitions between the $m_s = 0$ and $m_s = \pm 1$ states will thus lower the PL rate whenever the microwave frequency hits one of these two transitions. The resulting contrast in PL of roughly $10-15~\%$ \cite{Robledo2011}, indicated in figure \ref{fig:NVSpec_ODMR} (d), enables the optical detection of the electronic spin state. 
Furthermore, the frequency spacing between both dips is proportional to the projection of the magnetic field $B_z$ on the high symmetry axis of the NV, enabling it to act as a nanoscale magnetic field sensor. The sensitivity $\eta_\text{DC}$ of a single NV to a static magnetic field in case of the off-resonant readout method described here is given by the following equation \cite{Rondin2014}.
\begin{equation}
\eta_\text{DC} \approx \frac{h}{g \mu_\text{B}} \frac{\Delta \nu}{\sqrt{I_0} C}
\label{eq:SensNV}
\end{equation}
Here, $\Delta \nu$ is the linewidth of the ODMR resonance depicted in figure \ref{fig:NVSpec_ODMR} (d) and $C$ the corresponding contrast. Whereas the latter is limited by internal dynamics and microwave as well as laser power, the former is fundamentally-limited by the inverse of the coherence time $T_2^\star$. 
One way to tune the sensitivity is the parameter $I_0$, which is proportional to the detected PL rate from the NV. 
For $I_0$, the limiting factor is the comparably high refractive index of diamond ($n=2.41$ at 640 nm \cite{Phillip1964}), leading to a strong confinement of emitted PL inside diamond. Nanopillars guiding the PL towards the detection optics can overcome this limitation and simultaneously serve as a robust tip for scanning probe techniques.
\section{Simulation Setup}
For most of the simulations shown here, we use a commercial-grade simulator based on the finite-difference time-domain method (\textit{Lumerical FDTD-Solutions}). To ensure the reliability of our simulations, we performed comprehensive convergence tests. From these tests, we extract the numerical error, occurring mainly due to the finite size of the mesh cells and time steps in the simulations. For some illustrations, we additionally use a commercial-grade finite element simulator (\textit{COMSOL Multiphysics}).
\subsection{Implementation}
\begin{figure}[h]
\centering
\includegraphics[width=.7\linewidth]{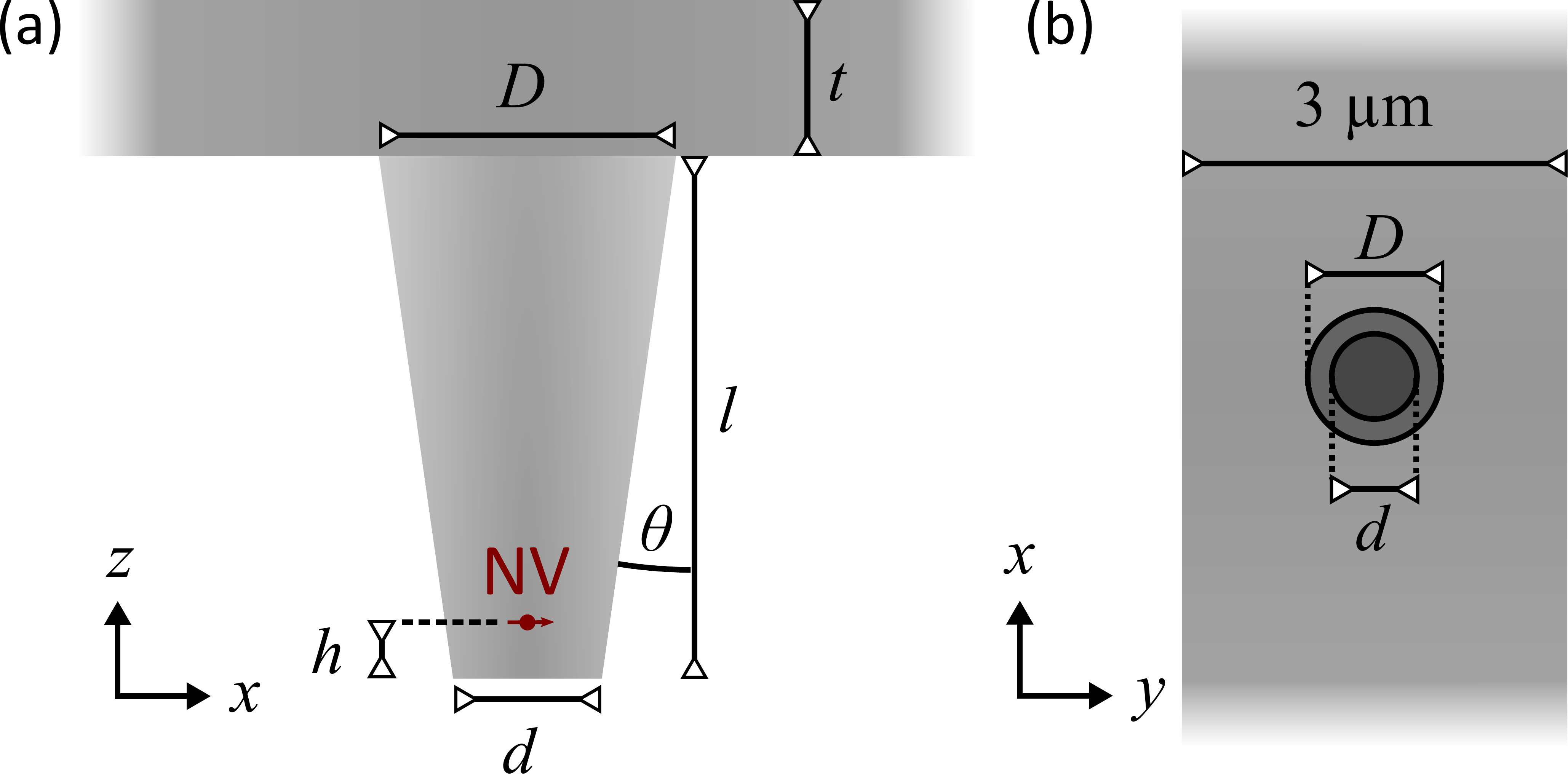}
\caption{Side view (a) and top view (b) of the simulated system. The taper angle $\theta$, the top diameter $d$ and the length $l$ of the pillar define the base diameter via $D =d + 2 l \tan(\theta)$. The NV is depicted as red arrow in a depth $h$ below the top facet of the pillar.}
\label{fig:SimulationSetup}
\end{figure}

Figure \ref{fig:SimulationSetup} sketches the general geometry of the scanning probe device for our simulations. The pillar with the NV inside is implemented as a tapered, truncated cone with length $l$, top diameter $d$ and taper angle $\theta$, see figure \ref{fig:SimulationSetup} (a). Note that assuming a truncated cone, instead of a cylinder as in previous work \cite{Babinec2010}, is motivated by findings from our own nanofabrication and by previous work \cite{Momenzadeh2015}.
The pillar is attached to a thin diamond platform, which is assumed to be infinitely-extended in x-direction and 3 $\mu$m in y-direction, see figure \ref{fig:SimulationSetup} (b). Its thickness $t$ is set to $0.5~\mu$m unless otherwise stated, motivated by the device mounting procedure, which has to allow for a mechanical breaking of the platform to attach the device to an AFM head \cite{Appel2016}. We model diamond as a dispersionless dielectric with $n = 2.41$.
\par 
The transitions forming ZPL and PSB are electric dipole transitions, mediated via two electric point dipoles, that are both perpendicular to each other and to the symmetry axis of the NV \cite{Epstein2005}.
Assuming a $\lbrace 100\rbrace$ surface of the diamond, these dipoles can contribute both perpendicular and parallel dipole components with respect to that surface. As we show in the first part of our results (section \ref{sec:xy-position}), the perpendicular dipole shows no significant contributions to the collectible PL rate. Consequently, we model the NV subsequently as a single electric point dipole, oriented parallel to the diamond surface, i.e.\ in the xy-plane in figure \ref{fig:SimulationSetup} (b).

To obtain a high spatial resolution in scanning probe sensing with NVs, they have to be placed shallowly below the surface (typically $h \approx 10$ nm). To resolve the position of these shallow NVs in the simulation correctly, we applied an appropriately-sized mesh around the dipole. The whole simulation region is surrounded by perfectly-matched layers (PML) to simulate infinitely-extended half spaces and avoid artificial reflections. The specific mesh size as well as PML parameters were determined from the convergence tests.
\subsection{Figure of Merit}
We investigate the influence of different device geometries on the collectible PL rate from the PSB, that contains around 96 \% of the NV PL and is hence used for spin-readout. 
For single photon emitters in nanophotonic structures, the collection efficiency $\eta$ is a commonly-used figure of merit, given in equation \eqref{eq:SingleCollEff}. It is generally defined as the rate of far field photons in a solid angle $\Gamma_\text{NA}^{(\lambda)}$ defined by the numerical aperture (NA) of the objective, divided by the total radiative decay rate $\Gamma_\text{rad}^{(\lambda)}$ of the emitter. 
\begin{equation}
\eta^{(\lambda)} = \frac{\Gamma_\text{NA}^{(\lambda)}}{\Gamma_\text{rad}^{(\lambda)}}
\label{eq:SingleCollEff}
\end{equation}
The superscript $\lambda$ indicates that these values are usually wavelength-depended.
Especially for applications using pulsed excitation, $\eta$ is an important figure, because it quantifies the probability of retrieving a collectible photon from the emitter in its environment after an excitation. Unity collection efficiency ($\eta = 1$) guarantees one collectible photon per excitation, assuming unity quantum efficiency. However, a high value of $\eta$ does not necessarily correspond to a high absolute photon rate $\Gamma_\text{NA}$ for cw excitation. As long as $\Gamma_\text{NA}$ is close to $\Gamma_\text{rad}$, $\eta$ remains close to unity, no matter the actual collectible photon rate $\Gamma_\text{NA}$. In situations where $\Gamma_\text{rad}$ is significantly decreased, i.e.\ if the radiative lifetime $\tau$ significantly increases due to a lowered local density of states (LDOS), $\Gamma_\text{NA}$ is also decreased, although $\eta$ may still be close to unity.

For an off-resonant cw readout as described in section \ref{sec:magnetometry}, however, a high collectible photon rate $\Gamma_\text{NA}$ is essential for an enhanced sensitivity $\eta_\text{DC}$. A better-suited expression, which quantifies this absolute photon rate, is given by the collection factor $\xi$ in equation \eqref{eq:SingleCollFac}:
\begin{equation}
\xi^{(\lambda)} = \frac{\Gamma_\text{NA}^{(\lambda)}}{\Gamma_\text{hom}^{(\lambda)}}
\label{eq:SingleCollFac}
\end{equation}
Here, $\Gamma_\text{hom}^{(\lambda)}$ is the radiative decay rate of an electric dipole inside a homogeneous medium, which is in our case diamond. The collection factor $\xi^{(\lambda)}$ thus simply defines a handy and comparable value for the absolute collectible photon rate $\Gamma_\text{NA}^{(\lambda)}$. 

As the simulations are purely electromagnetic and only provide classical optical powers and no quantum-mechanical photon rates, we need to emphasize the following relation:
\begin{equation}
\frac{\Gamma_\text{rad}^{(\lambda)}}{\Gamma_\text{hom}^{(\lambda)}} = \frac{P_\text{rad}^{(\lambda)}}{P_\text{hom}^{(\lambda)}}
\label{eq:classicQED}
\end{equation}
Equation \eqref{eq:classicQED} can be derived by the comparison of the classical Greens function of an electric point dipole and the quantum mechanical decay rate of an electric dipole transition \cite{PrinciplesOfNanoOptics}. It links both classical optical powers $P$ and quantum-mechanical rates $\Gamma$ and allows us to restrict our investigations to purely classical simulations and thereby to fractions of optical powers, always with the assumption of unity quantum efficiency.
\\
To take the broad emission in the PSB into account, it is useful to average $\xi$ over the PSB:
\begin{equation}
\overline{\xi} = \frac{1}{N} \sum_{k=1}^{N} \xi^{(\lambda_k)}
\label{eq:AvgCollFac}
\end{equation}
This average collection factor $\overline{\xi}$ will be the major figure of merit for all upcoming simulations. We sample the PSB at $N=7$ wavelengths (640 nm, 660 nm, 680 nm, 700 nm, 720 nm, 740 nm, 760 nm). For all these wavelengths, we found a relative error for $\xi^{(\lambda)}$ of $3~\%$ with our convergence tests. This error originates from the trade-off in accuracy and computational efforts. 

In our simulations, we compute $\overline{\xi}$ as follows: Because $\xi^{(\lambda)}$ is directly related to the electromagnetic power inside the solid angle of the NA, we can decompose it into two different factors. First, we have the fraction $T_\Uparrow^{(\lambda)}$ of the radiated power reaching the upper half space above the diamond platform. For $\textrm{NA} = 1$ ($90^\circ$ collection angle), this fraction would already define $\xi^{(\lambda)}$. For smaller NA, we introduce a second factor $T_\text{NA}^{(\lambda)}$, namely the fraction of the radiated power, that reaches the upper half space and which, in the far field, propagates within the solid angle defined by the NA. We will in the following always assume an NA of $0.8$.
Both factors can be derived from our FDTD simulations, and $\overline{\xi}$ will be calculated as follows:
\begin{equation}
\overline{\xi}  = \frac{1}{N} \sum_{k=1}^{N} \underbrace{T_\Uparrow^{(\lambda_k)} \cdot T_\text{NA}^{(\lambda_k)}}_{\xi^{(\lambda_k)}}
\label{eq:avgcollfac_FDTD}
\end{equation}
As nanophotonic structures such as the pillar influence the LDOS and thus the radiated power $P_\text{rad}$, we additionally define $LDOS^{(\lambda)}$, which quantifies the change in LDOS:
\begin{equation}
LDOS^{(\lambda)} = \frac{P_\text{rad}^{(\lambda)}}{P_\text{hom}^{(\lambda)}}
\end{equation}
For $LDOS^{(\lambda)}>1$, the LDOS is increased compared to bulk diamond, corresponding to a higher radiated power or decay rate, and vice versa for $LDOS^{(\lambda)} < 1$. 

We also define $\overline{LDOS}$, $\overline{T}_\Uparrow$ and $\overline{T}_\text{NA}$ equivalently to $\overline{\xi}$ in equation \eqref{eq:AvgCollFac}.
\section{Results}
As a first step, we show that only a dipole oriented in the xy-plane contributes significantly to $\overline{\xi}$. Next, we sweep the geometric parameters of the pillar and the platform and explain their influence on $\overline{\xi}$ and the underlying mechanisms. Finally, we optimize all parameters to maximize $\overline{\xi}$ and discuss the feasibility of the optimized device regarding our nanofabrication. To have a comparison value for the upcoming results, we calculate $\overline{\xi}=0.03$ for a single dipole in bulk diamond, oriented parallel to a surface, which defines a transition to air.

The default parameters in all upcoming simulations are, unless otherwise stated, $\theta = 0^\circ$, $d=D=200$ nm and $l=1~\mu$m (definitions cf.\ figure \ref{fig:SimulationSetup}), yielding a cylindrical pillar. The default thickness of the platform is $t = 0.5~\mu$m.
\subsection{Dipole Orientation \& Position}
\label{sec:xy-position}
To investigate the influence of the dipole orientation on $\overline{\xi}$, we simulate dipoles oriented along all three Cartesian axes defined in figure \ref{fig:SimulationSetup} at different lateral positions inside the pillar for $h=10$ nm. The simulation of x-, y- and z-dipoles allows to deduce $\overline{\xi}$ for an arbitrary dipole orientation. 
We perform these simulations for $d = 200$ nm and $d=300$ nm, motivated by the cylindrical pillar beeing a well-guiding optical fiber \cite{Bures2009}, which is single-mode for $d=200$ nm and multi-mode for $d=300$ nm for all wavelengths in the PSB. Because of the symmetry of the system, it is sufficient to simulate only one quarter of the pillar. 
\par
\begin{figure}[h]
\centering
\includegraphics[width=\linewidth]{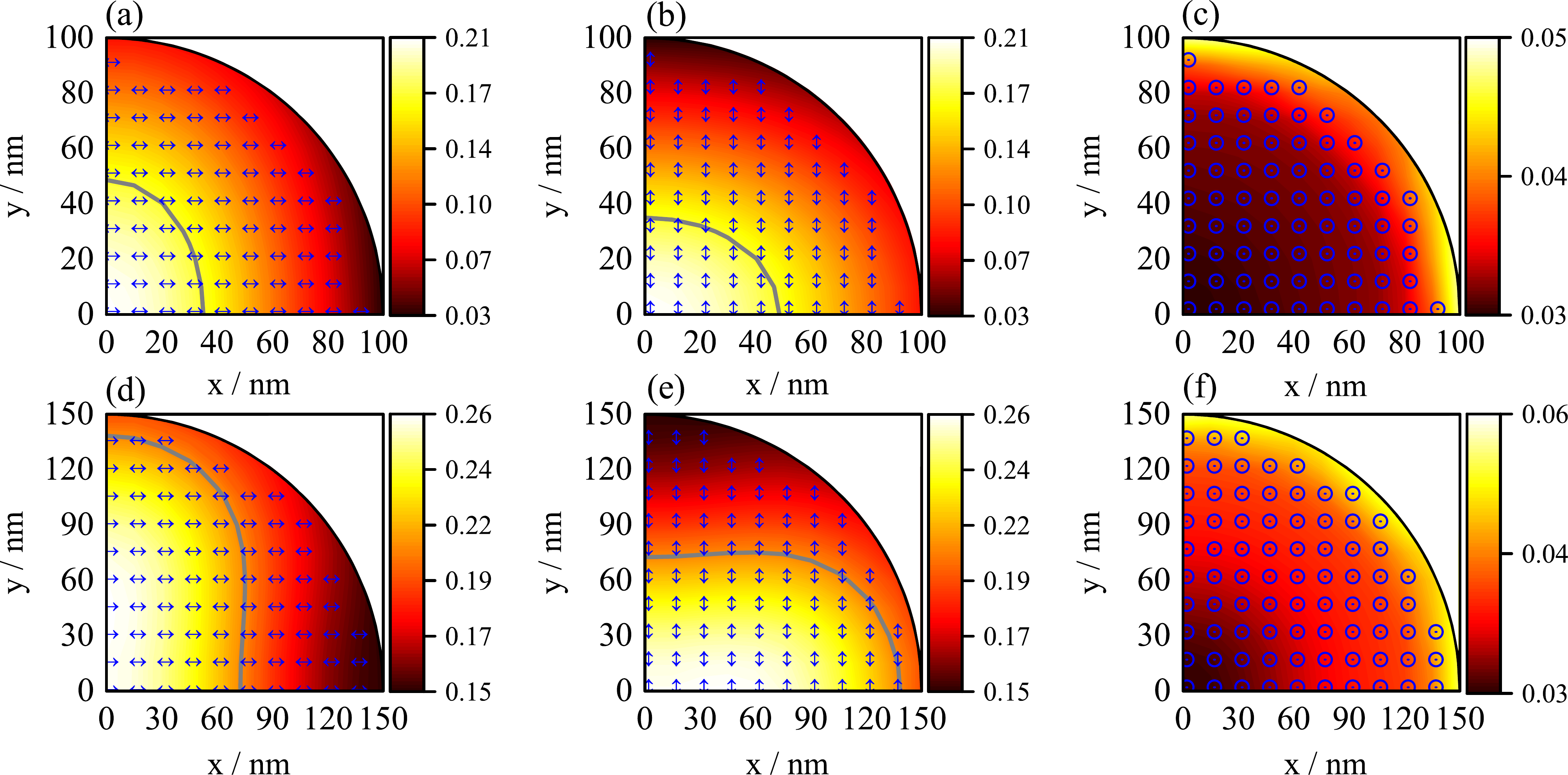}
\caption{Influence of different lateral dipole positions on $\overline{\xi}$ for a dipole at a depth of $h=10$ nm, oriented in $x$-, $y$- and $z$-direction, and a pillar with a diameter of $d=200$ nm (a),(b),(c) and $d=300$ nm (d),(e),(f). The circles ($z$-dipole), respectively arrows ($x$-/$y$-dipole), depict the actually-simulated positions and orientations, the contours in-between are interpolated. The gray contour line for the x- and y-dipole indicate $80~\%$ of the maximum $\overline{\xi}$ achieved in the center. The asymmetry of this contour originates from the asymmetric electric field distribution of the \HE modes, compare figure \ref{fig:ModeExpansion} (a).}
\label{fig:XY}
\end{figure}

Figure \ref{fig:XY} shows the resulting $\overline{\xi}$ for the different dipole orientations and positions. 
For the x- and y-dipole, shown in figure \ref{fig:XY} (a), (b), (d) and (e), $\overline{\xi}$ maximizes in the lateral center of the pillar, yielding $\overline{\xi}=0.20$ for $d=200$ nm and $\overline{\xi}=0.26$ for $d=300$ nm, whereas it minimizes here for the z-dipole, see figure \ref{fig:XY} (c) and (f).
\par
\begin{figure}[h]
\centering
\includegraphics[width=\linewidth]{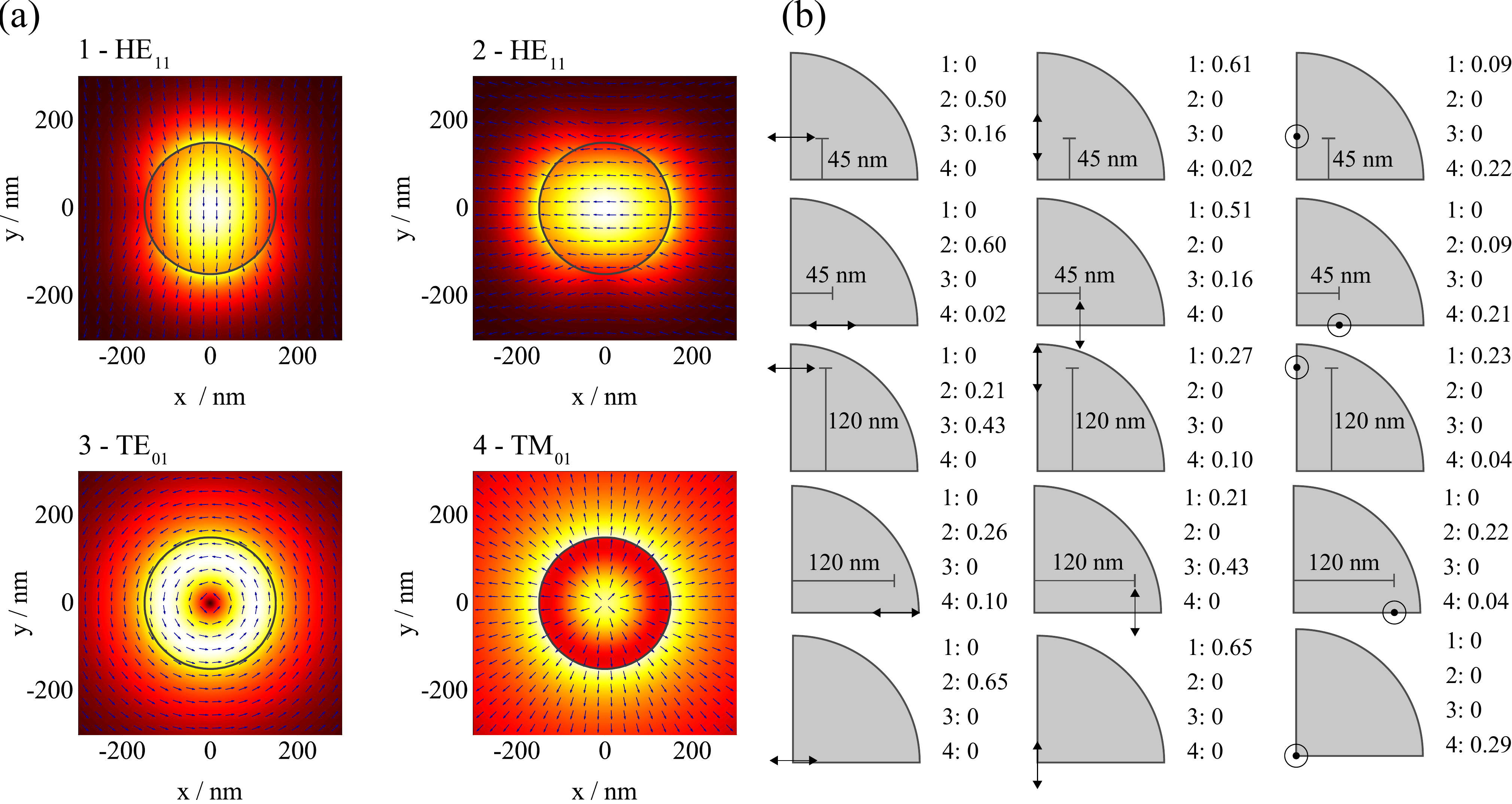}
\caption{For a pillar with $d=300$ nm, four guided modes exist (a). Shown are the squared amplitudes of the electric fields for $\lambda = 700$ nm. The arrows indicate the polarization of the mode projected on the xy-plane. For $d=200$ nm, only the two fundamental \HE modes exist, but with a weaker confinement compared to $d=300$ nm. Using a mode expansion approach, we analyzed the coupling of the dipole to these four modes for specific dipole positions and orientations (b). The arrows (xy-dipole), respectively circles (z-dipole), indicate the dipole orientation. The numbers 1 to 4 link to the four guided modes in (a), the numbers behind quantify the fraction of the total radiated power $P_\text{rad}^{(700~\text{nm})}$ that goes into this modes, usually called $\beta$-factor.}
\label{fig:ModeExpansion}
\end{figure}

Modelling the cylindrical pillar as a well-guiding optical fiber explains these results very well.
The dipole couples to the discrete set of guided and leaky modes of the pillar as well as to the continuum of radiation and evanescent modes. However, only the guided modes will be transmitted efficiently to the upper half space and contribute significantly to $\overline{\xi}$. As a consequence, increasing the coupling to the guided modes directly increases $\overline{\xi}$. According to the LDOS formalism, the radiative emission rate $\Gamma_\text{rad}$ into a certain mode is proportional to the overlap of the dipole moment $\textbf{d}$ and the electric field strength of this mode at the position of the dipole $\textbf{E}_{\textbf{k}}(\textbf{r}_0)$ \cite{PrinciplesOfNanoOptics}. 
\begin{equation}
\Gamma_\text{rad} \propto \sum_{\textbf{k}}  \left| \textbf{d} \cdot \textbf{E}_{\textbf{k}}(\textbf{r}_0) \right|^2 \delta\left(\omega_0 - \omega_\textbf{k} \right) 
\label{eq:LDOS}
\end{equation}
Because of the dot product in equation \eqref{eq:LDOS}, the polarization of the mode needs to be aligned to the orientation $\textbf{d}$ of the dipole to achieve a significant coupling.
The two fundamental guided modes of the pillar named \HE are orthogonally-polarized to each other within the xy-plane, see figure \ref{fig:ModeExpansion} (a). These so-called hybrid modes possess both electric as well as magnetic field components in direction of propagation, yielding a hybrid polarization: In the center of the pillar, they are dominantly-polarized within the xy-plane, whereas the fraction of z-polarization raises towards the edges of the pillar. In addition, their electric field strength peaks in the lateral center of the pillar, maximizing $\overline{\xi}$ for an x-/y-dipole placed here. For the same reason, a centered z-dipole does not couple to the \HE modes, consistent with previous work \cite{Sondergaard2001}.
\par
For $d=300$ nm, the pillar is multi-mode for all wavelengths in the PSB. Additionally, the confinement of the \HE modes increases, yielding a higher field strength inside the pillar and thus a higher LDOS and $\overline{\xi}$. We will discuss this further in section \ref{sec:length}, where we sweep the diameter of the pillar.
The two additional guided modes, called \TE and \TM mode, possess both circular symmetry. Whereas the \TE mode is polarized within the xy-plane, the \TM mode has a non-vanishing electric field component in z-direction, which may result in a higher LDOS for the z-dipole. Using a mode expansion approach, we evaluate the coupling of selected dipole configurations to these modes, see figure \ref{fig:ModeExpansion} (b). The given numbers are the $\beta$-factors, which describe the fraction of total power $P_\text{rad}^{(700~\text{nm})}$ radiated into a specific mode.
As suspected, the centered z-dipole significantly couples to the \TM mode ($\beta = 0.29$), for which $\textbf{E}_{\textbf{k}}$ peaks in the center of the pillar. The mode expansion also reveals a rising coupling of the z-dipole to the fundamental \HE modes towards the edge of the pillar, what we attribute to their hybrid character. Yet the contribution of the z-dipole to $\overline{\xi}$ via both modes remains negligibly low, see figure \ref{fig:XY} (f). In conclusion, placing the NV in the lateral center of the pillar maximizes $\overline{\xi}$. Since the x-/y-dipoles dominate the contribution to $\overline{\xi}$, we neglect the z-dipole in the following.

Placing NVs with nanometer accuracy is challenging, but can be achieved using nanoimplantation techniques involving a pierced AFM tip \cite{Meijer2008,Spinicelli2011}. However, to ease the fabrication of a large number of devices, nitrogen ions are often implanted homogeneously into the diamond and the pillars are etched subsequently \cite{Appel2016}. In this case, the implanted NVs are statistically distributed in the xy-plane.
Notably, our simulations reveal a plateau of high $\overline{\xi}$ for the x- and y-dipole around the center of the pillar. The gray contour lines in figure \ref{fig:XY} enclose an area, where $\overline{\xi}$ stays above $80~\%$ of its maximum value in the center. The NV thus does not need to be placed exactly in the center, but can have a radial offset with only a small reduction of $\overline{\xi}$, increasing the resulting yield of devices with acceptable $\overline{\xi}$.
\par
In contrast, we found the depth $h$ of the dipole to negligibly influence $\overline{\xi}$. Especially in the context of sensing, $h$ is a crucial parameter, fundamentally limiting the achievable spatial resolution in sensing and imaging applications. It is thus favorable to position the NV as shallowly as possible. For very shallow NVs, however, a significant broadening of their electron spin resonance was found \cite{Ofori-Okai2012}, limiting the minimal depth to a reasonable value of $h=2$ nm. Simulating a laterally-centered x-dipole with $h = 2$ nm to $30$ nm within our default configuration ($d=200$ nm, $l=1~\mu$m, $\theta=0^\circ$ and $t=0.5~\mu$m) results in a maximum $\overline{\xi}$ at $h=13$ nm. However, the variation in $\overline{\xi}$ over the investigated range of $h$ is less than the numerical error ($3~\%$). According to SRIM (Stopping and Range of Ions in Matter) simulations, a typical straggle for such implantation depths is in the order of $4$ nm (7 keV implantation energy, yielding a depth of $h=12$ nm). Consequently, $\overline{\xi}$ does not significantly change for NVs created within the straggle of the implantation. The resulting yield of usable devices should therefore not be limited by the vertical, but only by the lateral positioning accuracy.
\subsection{Pillar Geometry}
\label{sec:length}
\begin{figure}[h]
\centering
\includegraphics[width=\linewidth]{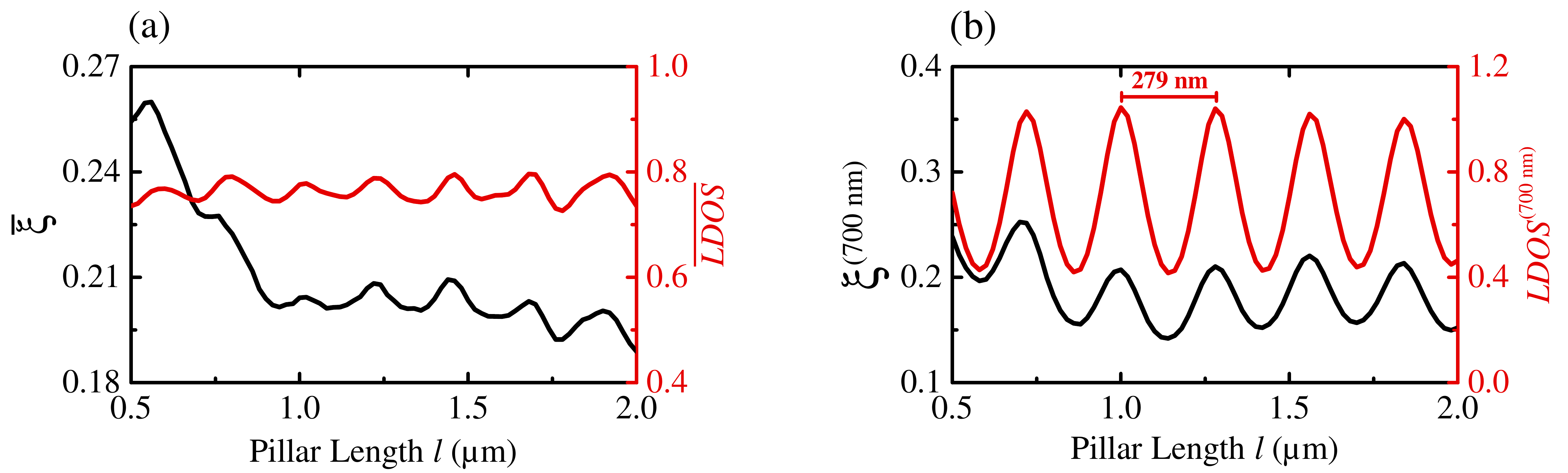}
\caption{Influence of the length $l$ of a cylindrical pillar with $d=200$ nm on $\overline{\xi}$ and $\overline{LDOS}$ (a), and on $\xi^{(700~\text{nm})}$ and $LDOS^{(700~\text{nm})}$ (b). We attribute the oscillations to standing waves inside the pillar that form a node or an anti-node at the position of the dipole, changing the LDOS and thus $\overline{\xi}$ depending on $l$. The oscillations of $\xi^{(700~\text{nm})}$ have a period of 279 nm, equal to half the material wavelength of the \HE modes. The rise in $\overline{\xi}$ for shorter pillars cannot be explained by $\overline{LDOS}$, but with an enhanced far field confinement $T_\text{NA}$ for $l<0.9~\mu$m.}
\label{fig:Length}
\end{figure}

First, we investigate the influence of the pillar length $l$ on $\overline{\xi}$ for the default configuration, which includes now a laterally-centered x-dipole at $h=10$ nm. 
We obtain a maximum $\overline{\xi} = 0.26$ for $l = 0.55~\mu$m, see figure \ref{fig:Length} (a). Towards $l=0.9~\mu$m, $\overline{\xi}$ drops to $0.2$. For $l>0.9~\mu$m, an oscillatory behaviour emerges, which we attribute to Fabry-Pérot resonances, i.e. standing waves inside the pillar: The guided modes define the wavelength-depended effective mode index $n_\text{eff}$ and thus the optical length $n_\text{eff}\cdot l$ of the pillar.
Depending on that optical length, the standing waves form a node or an anti-node at the position of the dipole, altering the LDOS and thereby the coupling to the corresponding guided mode. 
Consequently, $\overline{LDOS}$ also shows oscillations for varying $l$, which correlate to those of $\overline{\xi}$. These are even more pronounced for a single wavelength, e.g. $\lambda = 700$ nm, shown in figure \ref{fig:Length} (b). Here, the oscillation periods of $\xi^{(700~\text{nm})}$ and $LDOS^{(700~\text{nm})}$ are perfectly correlated and fit to half the material wavelength, considering the effective mode index $n_\text{eff} = 1.257$ of the \HE modes for $d=200$ nm and $\lambda = 700$ nm.
For $\overline{\xi}$, this effect is blurred, because of the wavelength-dependence of $n_\text{eff}$ (e.g. $n_\text{eff}=1.407$ for $\lambda=640$ nm and $n_\text{eff}=1.144$ for $\lambda=760$ nm), changing the optical length of the pillar for the different wavelengths within the PSB. The amplitude of the oscillations in $\overline{\xi}$ reduces to around $3~\%$, which is the same order as the numerical error. Notably, the higher $\overline{\xi}$ for short pillars is not a consequence of an increased LDOS, but of an enhanced far field confinement $\overline{T}_\text{NA}$.
\par
In conclusion, we found only a weak dependence of $\overline{\xi}$ on $l$ when collecting the whole PSB of the NVs. For $l<0.9~\mu$m, there is a substantial increase in $\overline{\xi}$, yet such very short pillars might not be usable to scan all types of samples.
\par
\begin{figure}[h]
\centering
\includegraphics[width=\linewidth]{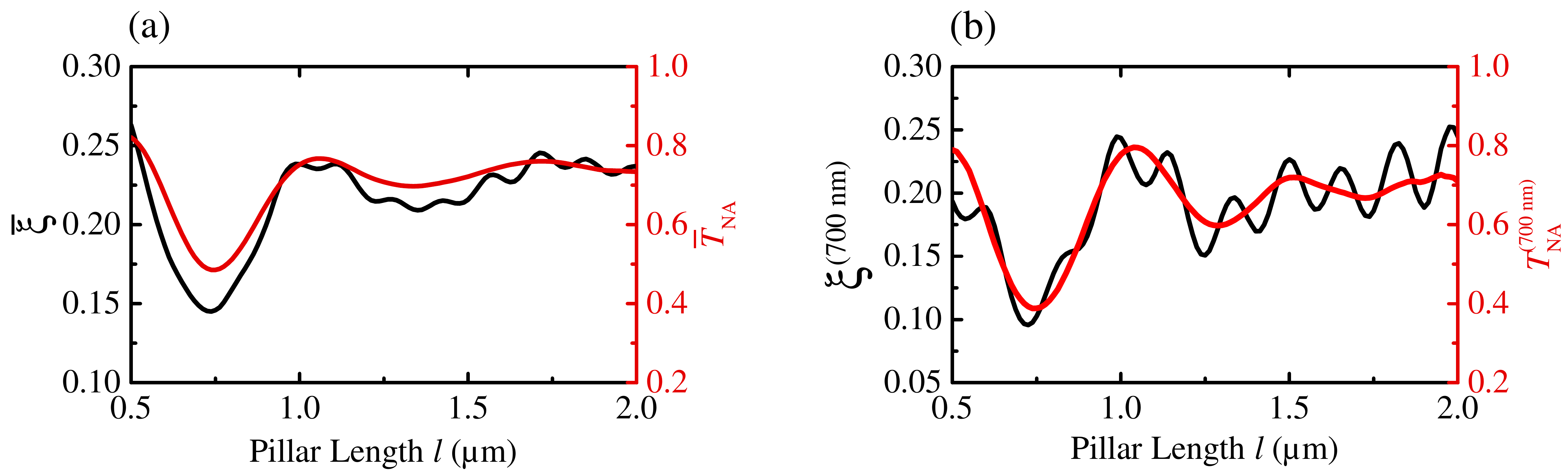}
\caption{Influence of the length $l$ of a tapered pillar ($\theta = 4^\circ$, $l=1~\mu$m, $d=200$ nm) on $\overline{\xi}$ and $\overline{T}_\text{NA}$ (a). The oscillations in $\overline{\xi}$ still correlate to $\overline{LDOS}$, but the dip in collection factor for shorter pillars is a result of a bad far field confinement $\overline{T}_\text{NA}$. This can also be seen from $\xi^{(700~\text{nm})}$ and $T_\text{NA}^{(700~\text{nm})}$ (b), where the oscillations are even more pronounced.}
\label{fig:Length_4deg}
\end{figure}

The cylindrical pillar assumed to this point provides insights into the mechanisms behind the coupling of the dipole to the guided modes of the pillar.  
However, with our nanofabrication, where we use inductively-coupled plasma reactive ion etching (ICP-RIE\footnote{Oxford PlasmaLab 100: 50 sccm Ar, 50 sccm O$_2$, 200 W RF power, 500 W ICP power, 465 V bias voltage}), we usually obtain slightly tapered pillars. From scanning electron microscopy (SEM, Hitachi S800) images, we determine taper angles between $3^\circ$ and $5^\circ$ for pillars etched using the same plasma conditions. For $\theta > 0^\circ$, the diameter of the pillar linearly increases from $d$ at the top to $D=d+2\cdot l \cdot \tan(\theta)$ at the base. Figure \ref{fig:Length_4deg} shows again $\overline{\xi}$ for varying $l$, but this time with $\theta = 4^\circ$ (giving $D=340$ nm for $d=200$ nm).
\par
The resulting curve shows two superimposed oscillations in $\overline{\xi}$: One with small amplitudes and a short period, originating from LDOS oscillations comparable to the case with $\theta = 0^\circ$ in figure \ref{fig:Length}, and one with comparably large amplitudes and a long period, what we attribute to the far field confinement, see the red lines in figure \ref{fig:Length_4deg}. For $\theta = 0^\circ$, we found it to be important for short pillar lengths with $l < 0.9~\mu$m, here its influence on $\overline{\xi}$ is also significant for longer pillars. Notably, $\overline{\xi}$ in average increased for $l>0.9~\mu$m compared to the untapered pillar, in this case making tapered pillars favorable over cylindrical pillars.
\par
\begin{figure}[h]
\centering
\includegraphics[width=\linewidth]{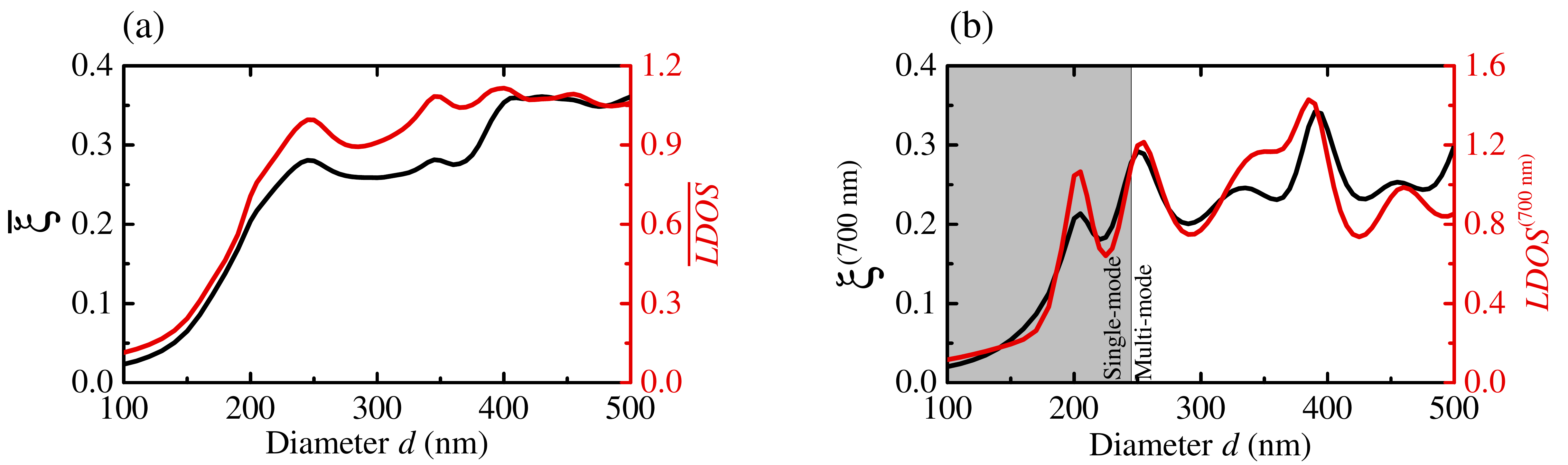}
\caption{Influence of the top diameter $d$ of a cylindrical pillar with $l=1~\mu$m on $\overline{\xi}$ and $\overline{LDOS}$ (a), and on $\xi^{(700~\text{nm})}$ and $LDOS^{(700~\text{nm})}$ (b). For both cases, a similar behaviour of $\overline{\xi}$ and $\overline{LDOS}$, respectively $\xi^{(700 \text{nm})}$ and $LDOS^{(700 \text{nm})}$, exists: For $d<200$ nm, the weak mode confinement reduces the LDOS and thereby the coupling to the \HE modes. For $d>200$ nm, the confinement converges to unity, but the continuously increasing $n_\text{eff}$ leads again to pronounced Fabry-Pérot resonances for single wavelengths, reducing strongly when averaging over the PSB.} 
\label{fig:Diameter}
\end{figure}

The second important geometric parameter is the diameter of the pillar, because it defines $n_\text{eff}$ for the guided modes and thus the standing waves. We start again with our default configuration ($\theta = 0^\circ$, pillar length $l = 1~\mu$m, platform thickness $t = 0.5~\mu$m) and simulate varying $d$. The results are shown in figure \ref{fig:Diameter} (a), where we see a first maximum at $d=245$ nm, yielding $\overline{\xi} = 0.28$.
The monotone rise from $d=100$ nm to this maximum is due to the increasing confinement of the \HE modes: For $d < 200$ nm, the electric field strength of the modes concentrates mostly in the evanescent field, yielding low electric field strengths $\textbf{E}_{\textbf{k}}(\textbf{r}_0)$ in the lateral center of the pillar. 
\par
\begin{figure}[h]
\centering
\includegraphics[width=\linewidth]{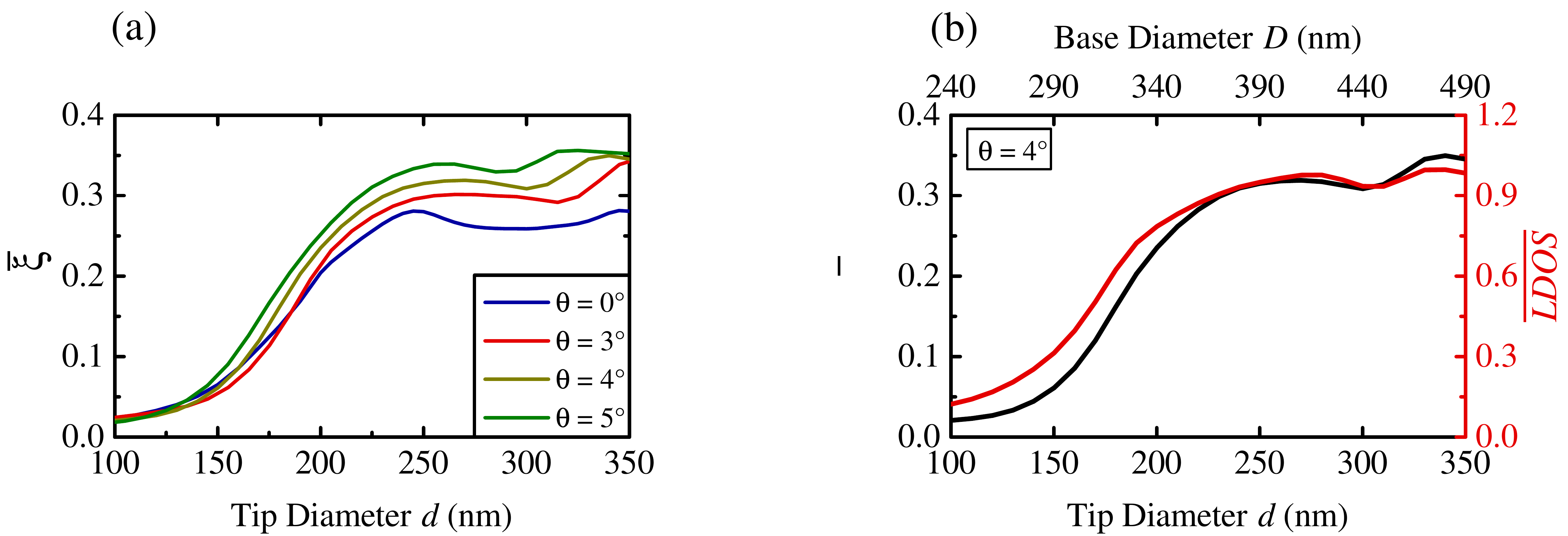}
\caption{Both figures show $\overline{\xi}$ for tapered pillars with $l=1~\mu$m. Comparing different taper angles (a), $\overline{\xi}$ behaves comparably to the case with $\theta = 0°$, but increasing $\theta$ shifts the first maximum to higher top diameters $d$ and also increases $\overline{\xi}$ in general. Looking at the case of $\theta = 4^\circ$ in detail (b), $\overline{\xi}$ (black) still follows $\overline{LDOS}$ (red), reaching a maximum value of $0.32$ at a top diameter $d=270$ nm.}
\label{fig:Tapered}
\end{figure}

For $d>245$ nm, the pillar becomes multi-mode for $\lambda = 700$ nm, yet this transition does not mark a significant feature for $\xi^{(700 \text{nm})}$, see figure \ref{fig:Diameter} (b), because the x-dipole does not couple to higher modes, as discussed in section \ref{sec:xy-position}. The oscillations of $\xi^{(700 \text{nm})}$ are a result of the continuously increasing effective mode index $n_\text{eff}$ of the \HE fundamental modes, leading again to Fabry-Pérot resonances. These oscillations again strongly reduce for $\overline{\xi}$. Consistently, the totally radiated power $P_\text{rad}^{(\lambda)}$ converges to the bulk power $P_\text{hom}^{(\lambda)}$ for increasing $d$, meaning that $\overline{LDOS}$ tends to unity for large $d$.
Introducing again a non-vanishing taper angle leads to the results in figure \ref{fig:Tapered}. 
\par
For $\theta = 0^\circ$, we found the first maximum $\overline{\xi} = 0.28$ at $D = d = 245$ nm. For $\theta = 4^\circ$, our most-probable taper angle in nanofabrication, this maximum increases to $\overline{\xi} = 0.32$ and occurs for a top diameter of $d = 270$ nm, corresponding to a base diameter of $D = 410$ nm with $l=1~\mu$m, see figure \ref{fig:Tapered} (b). Additionally, the width of this first maximum becomes broader for increasing $\theta$, potentially allowing larger fabrication tolerances.
\par
To sum up, this investigation of the geometric parameters of the pillar shows that the Fabry-Pérot resonances are the most dominant influence on $\overline{\xi}$ for untapered pillars. For the more realistic case of $\theta > 0^\circ$, however, we also found the resulting far field confinement to be important especially for short pillars with $l<1~\mu$m. Furthermore, a tapered pillar generally increases $\overline{\xi}$, and even bigger taper angles $\theta > 5^\circ$ could lead to a further increased $\overline{\xi}$, but would require a major change in our etching recipe.
\subsection{Platform Thickness}
\label{sec:platform}
\begin{figure}[h]
\centering
\includegraphics[width=\linewidth]{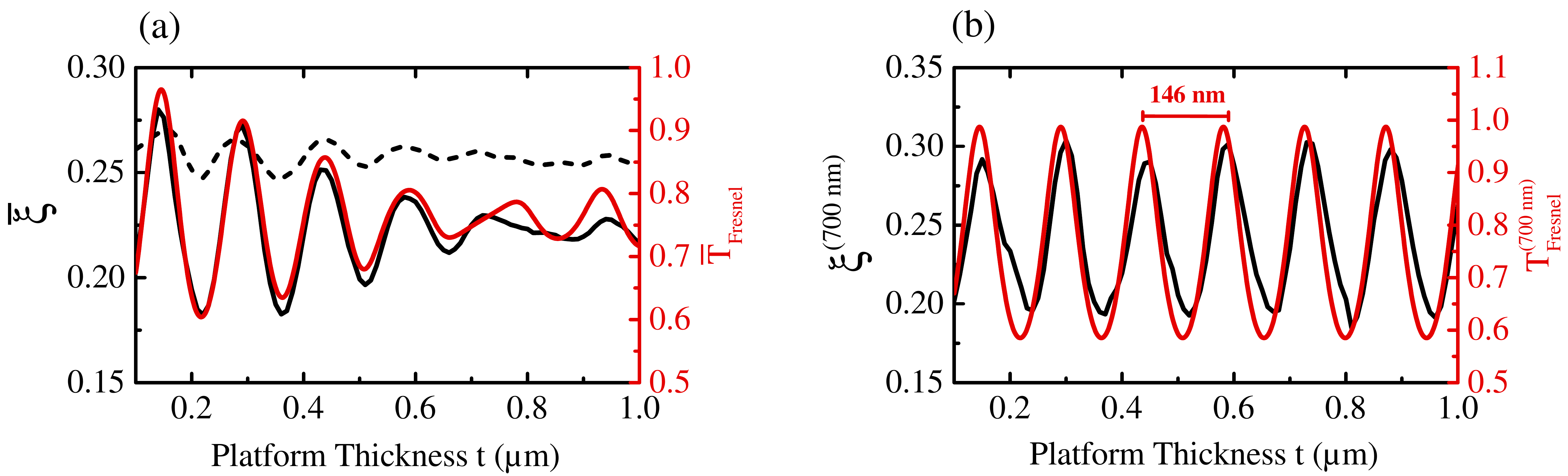}
\caption{Influence of the platform thickness $t$ for a cylindrical pillar ($d=200$ nm, $l = 1~\mu$m) on $\overline{\xi}$ (a) and on $\xi^{(700~\text{nm})}$ (b). For both cases, the Fresnel transmission $\overline{T}_\text{Fresnel}$ (a) and $T^{(700~\text{nm})}_\text{Fresnel}$ (b), respectively, are shown. To calculate these transmissions, we assume a plane wave incident on the platform, with a wave vector defined by the effective mode index of the \HE modes of the pillar. Modelling the platform as single dielectric layer then explains the observed oscillations: For $\lambda = 700$ nm, the oscillation period of 146 nm matches the expected half material wavelength for bulk diamond $700$ nm$/2.41/2$ quite well.}
\label{fig:Membrane}
\end{figure}

Our scanning probe device consists not only of the pillar, but it is placed on a thin ($< 1~\mu$m) diamond platform. The thickness of the platform $t$ is not crucial for the functionality of the device, given that it is thin enough to allow for mechanically detaching individual devices from the sample and attaching them to an AFM head \cite{Appel2016}, and thick enough to ensure mechanical stability (roughly $t > 0.2~\mu$m). However, the platform influences the photonic properties of the device as summarized in figure \ref{fig:Membrane}.
For $t < 0.7~\mu$m, $\overline{\xi}$ oscillates with a high amplitude of up to $30~\%$ of its mean value, but these oscillations damp strongly for $t > 0.7~\mu$m. Here, in contrast to the oscillations for different pillar dimensions shown in section \ref{sec:length}, they do not correlate to $\overline{LDOS}$, indicating another mechanism causing the oscillations.
We derive a simple explanation by considering the platform independently: The \HE modes of the pillar can be approximated as plane waves with a wave vector $\textbf{k} = k_0 \cdot n_\text{eff} \cdot \textbf{e}_z$, propagating through the platform towards the upper half space.
Using Fresnel equations, we calculate the transmission $T_\text{Fresnel}$ through the platform, depending on its thickness $t$. This rather simplified model describes the oscillations of $\overline{\xi}$ quite well, see the red solid lines in figure \ref{fig:Membrane}.
\par
Notably, the oscillation amplitude for $\xi^{(700 \text{nm})}$ and $T^{(700 \text{nm})}_\text{Fresnel}$ is constant for all simulated $t$, but damps for higher platform thickness when considering $\overline{\xi}$ and $\overline{T}_\text{Fresnel}$. This is a result of averaging over the broad PSB, as the oscillations all possess slightly different periods for the different wavelengths (e.\ g. 136 nm for $\lambda = 640$ nm and 160 nm for $\lambda = 760$ nm), yielding a beating-like behaviour for $\overline{\xi}$ and $\overline{T}_\text{Fresnel}$.
Having identified the platform to act simply as a dielectric layer, previously demonstrated anti-reflective (AR) coatings with silica ($n = 1.46$) could further boost the transmission independently of the thickness of the platform \cite{Yeung2012}. Figure \ref{fig:Membrane} (a) shows the possible increase of $\overline{\xi}$ as a dashed line, yielding only weak residual oscillations with varying platform thickness. This might be beneficial for nanofabrication, what we will discuss further in section \ref{sec:Optimization}.
\subsection{Trenches}
\begin{figure}[h]
\centering
\includegraphics[width=\linewidth]{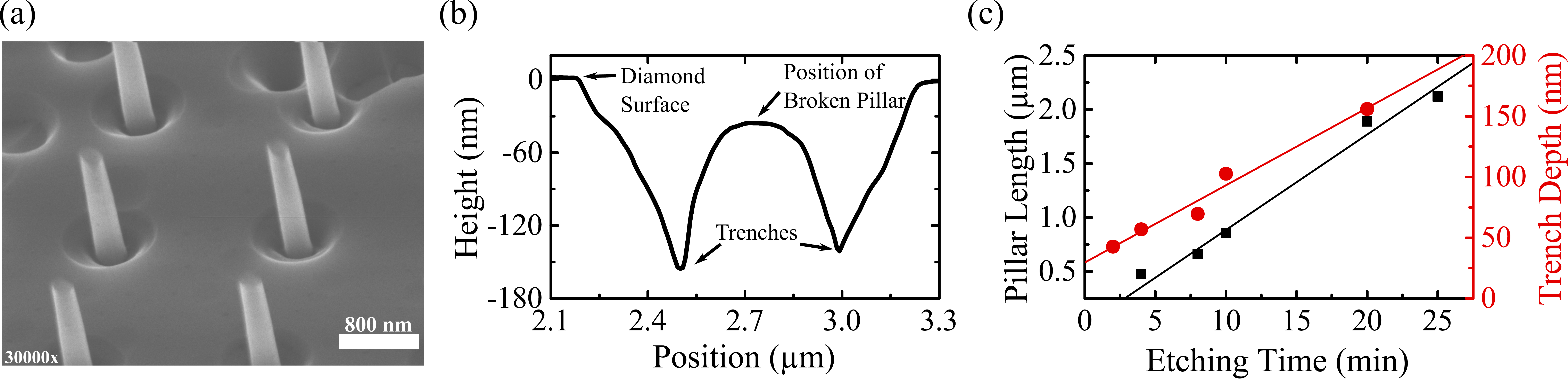}
\caption{An electron microscopy image of our fabricated pillars (a), here with a length of around $l=1.9~\mu$m and a taper angle of roughly $4^\circ$, clearly reveals the trenches around each pillar. We performed AFM measurements of samples where we remove the pillars to determine the actual trench geometry for different pillar lengths. An example is given in (b), where the AFM scan around the position of a pillar from part (a) is shown, yielding a mean width of 300 nm and depth of around 150 nm. Repeated measurements for pillars with different lengths reveal a linearly increasing trench depth with increasing etching time and thus with the pillar length (c).}
\label{fig:System_Trenches}
\end{figure}

During our ICP-RIE based nanofabrication, reflections of ions on the sidewalls of the evolving pillars are unavoidable. These reflected ions locally enhance the etching rate close to the pillar base and trenches form in the platform around the pillar \cite{Hoekstra1998,Challier2018}. 
Figure \ref{fig:System_Trenches} (a) shows an SEM image of our pillars with clearly visibile trenches. To determine the geometry of the trenches, we investigate pillars fabricated with different lengths from $l = 0.5~\mu$m to $l=2~\mu$m. Breaking the pillars in an ultrasonic bath allows us to measure the trench geometry with an AFM (Bruker FastScan, tapping mode), figure \ref{fig:System_Trenches} (b) shows an example of a corresponding scan. Repeating these scans for several pillars with different lengths, we find the trench depth to be proportional to the etching time, see figure \ref{fig:System_Trenches} (c). The trenches evolve with a rate of $6.4(6)$ nm/min. Additionally, we determine the pillar lengths via SEM imaging, yielding an etching rate of $88(3)$ nm/min. 
In contrast to that, we found no correlation between the trench width and the etching time, but the average trench width we see is between 100 and 300 nm. 

As the trenches are filled with air, they extend the length $l$ of the pillar and reduce the thickness $t$ of the platform. Based on the analysis we did in section \ref{sec:length} and \ref{sec:platform}, we would expect $\overline{\xi}$ to oscillate when sweeping the trench depth. As $\overline{\xi}$ oscillates stronger with varying $t$ than with varying $l$ (compare figures \ref{fig:Length_4deg} (a) and \ref{fig:Membrane} (a)), we expect the change in $t$ to be the dominant effect arising from the trench. 
\par
\begin{figure}[h]
\centering
\includegraphics[width=.8\linewidth]{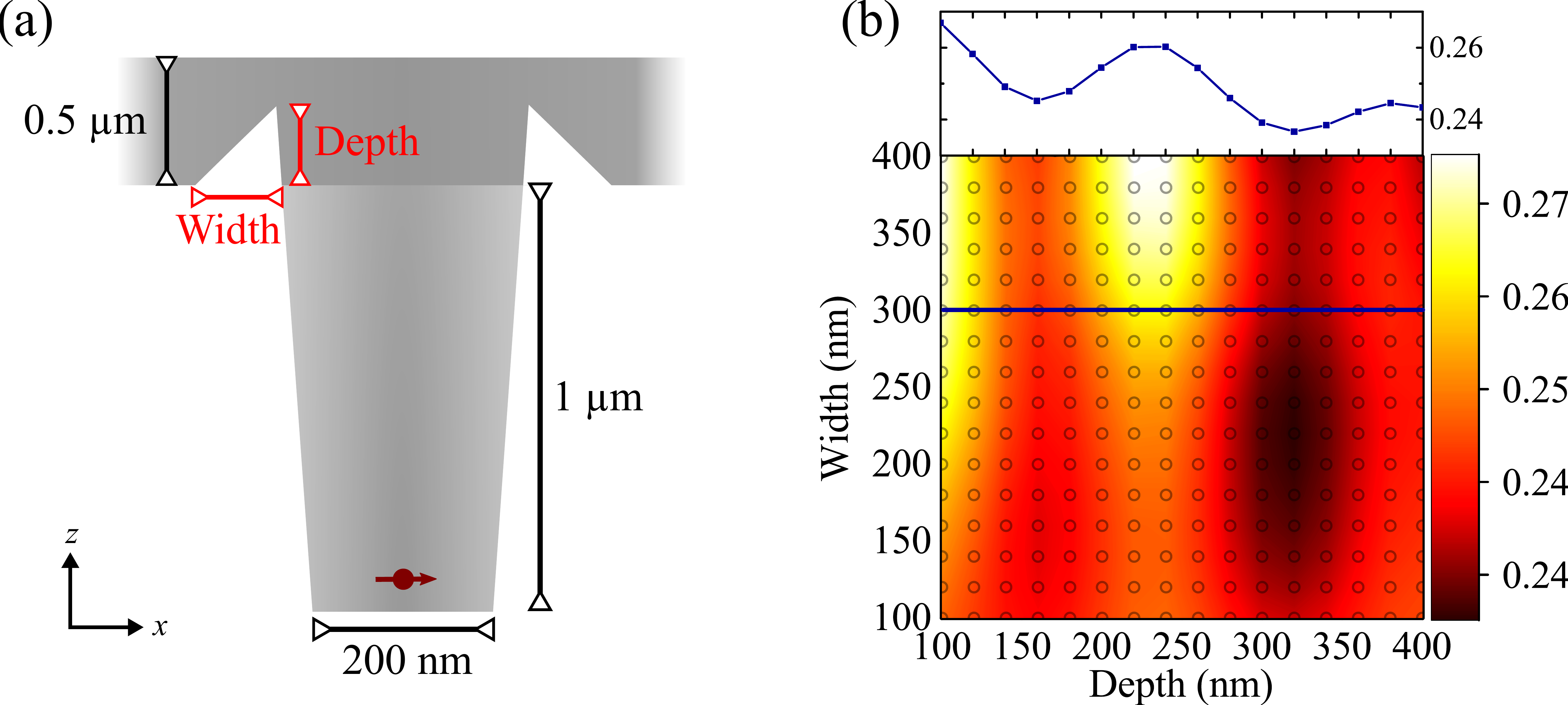}
\caption{We implement the trenches as shown in (a) for a device geometry with $t = 0.5~\mu$m, $l = 1~\mu$m, $d=200$ nm and $\theta = 4^\circ$. Sweeping the width and depth of the trench yields the results shown in (b), revealing the trenches to have a rather weak influence on $\overline{\xi}$. Looking at a specific trench width (dark blue line), varying the trench depth leads again to oscillations due to the decreasing residual platform thickness and simultaneously increasing effective pillar length.}
\label{fig:System_Trenches_2}
\end{figure}

To study the actual influence of the trench, we implemented it in our simulations as sketched in figure \ref{fig:System_Trenches_2} (a). The trench depth evolves linearly with the trench width. Sweeping its width and depth while all other parameters remain fixed ($d=200$ nm, $l=1~\mu$m, $t=0.5~\mu$m, $\theta = 4^\circ$) yields the results shown in figure \ref{fig:System_Trenches_2} (b).
\par
The blue line in figure \ref{fig:System_Trenches_2} (b) is a profile line trough the contour plot at a trench width of 300 nm, where the expected oscillations for varying trench depth are cleary visible, in accordance with the simple model of a trench increasing $l$ and decreasing $t$. If the trenches become too narrow, this effect vanishes and the amplitudes of the oscillations reduce.
Looking at the results in general, we can conclude that in the majority of all cases shown here, especially in the experimentally-found range of possible trench depths, a trench has a rather negligible influence on $\overline{\xi}$.
\section{Joint Optimization}
\label{sec:Optimization}
All parameters presented previously influence $\overline{\xi}$, however, it is possible to optimize the pillar independently of the plaform: The diameter of the pillar defines the confinement and effective index of the guided modes. The former restricts us to pillars with $d>200$ nm, otherwise the bad confinement reduces the LDOS and thus $\overline{\xi}$ drastically. The latter determines together with the length $l$ and taper angle $\theta$ of the pillar the guided modes. Tuning $d$ and $l$ should therefore be done first, followed by the thickness $t$ of the platform. In theory, one could additionally tune the taper angle $\theta$ prior to that. In practice, however, it is necessary to take the limits of nanofabrication into account.

ICP-RIE transfers our device geometry from our etching mask (FOx-16, Dow Corning), structured by electron beam lithography (EBL, Hitachi S 4500), to the diamond substrate. 
Perfect conditions, i.e.\ perfectly anisotropic etching with same conditions everywhere on the sample as well as vanishing mask erosion, should yield cylindrical pillars with $\theta \approx 0^\circ$.
The residual taper angle we see is most probably a result of a complex process involving both a nonideal plasma as well as significant mask erosion: Diamond is an insulator and thus experiences charging effects during the plasma etching. Randomly-varying local charge densities may lead to inhomogeneities in the plasma and a reduced anisotropy. Also mask erosion is known to cause tapered sidewalls. In previous work, the EBL exposure dose for the etch mask has been tuned together with the plasma parameters of the subsequent etching to introduce tapered sidewalls on purpose \cite{Momenzadeh2015}.
In our case, a nonideal EBL might also produce non-perfectly cylindrical masks. Erosion of these conical masks can hence lead to a decreasing mask diameter during the etching, yielding a truncated cone instead of a cylinder. Because of these complex causes of the residual taper angle, we do presently not consider it feasible to set $\theta$ experimentally to other values than the $3^\circ-5^\circ$ we achieve. However, we want to emphasize that we found this range to be constant over several batches. Consequently, we start our optimization with a fixed taper angle of $\theta = 4^\circ$, representing the most probable value with our current etching recipe, for which we can extract our optimal top diameter of $d=270$ nm from figure \ref{fig:Diameter}. Note that also the used implantation dose has to be adjusted to yield an NV density corresponding to a single NV on average in the area defined by the diameter of the pillar.
Starting from this, we swept first $l$, followed by $t$, to find a device geometry that maximizes $\overline{\xi}$.

We found a maximum $\overline{\xi} = 0.41$ for $l=1.65~\mu$m and $t = 0.43~\mu$m. Together with $\theta = 4^\circ$ and $d = 270$ nm, this declares our optimized device geometry, defining also the base diameter to $D = 501$ nm. To have a comparison value, we can extract $\overline{\xi}=0.23$ for the device geometry we first realized ($l=2~\mu$m, $d=200$ nm, $\theta=4^\circ$ and $t=0.5~\mu$m). This translates to around $1.8\times$ enhancement of $\overline{\xi}$ for our optimized device geometry compared to our current, non-optimized geometry.

\begin{figure}[h]
\centering
\includegraphics[width=\linewidth]{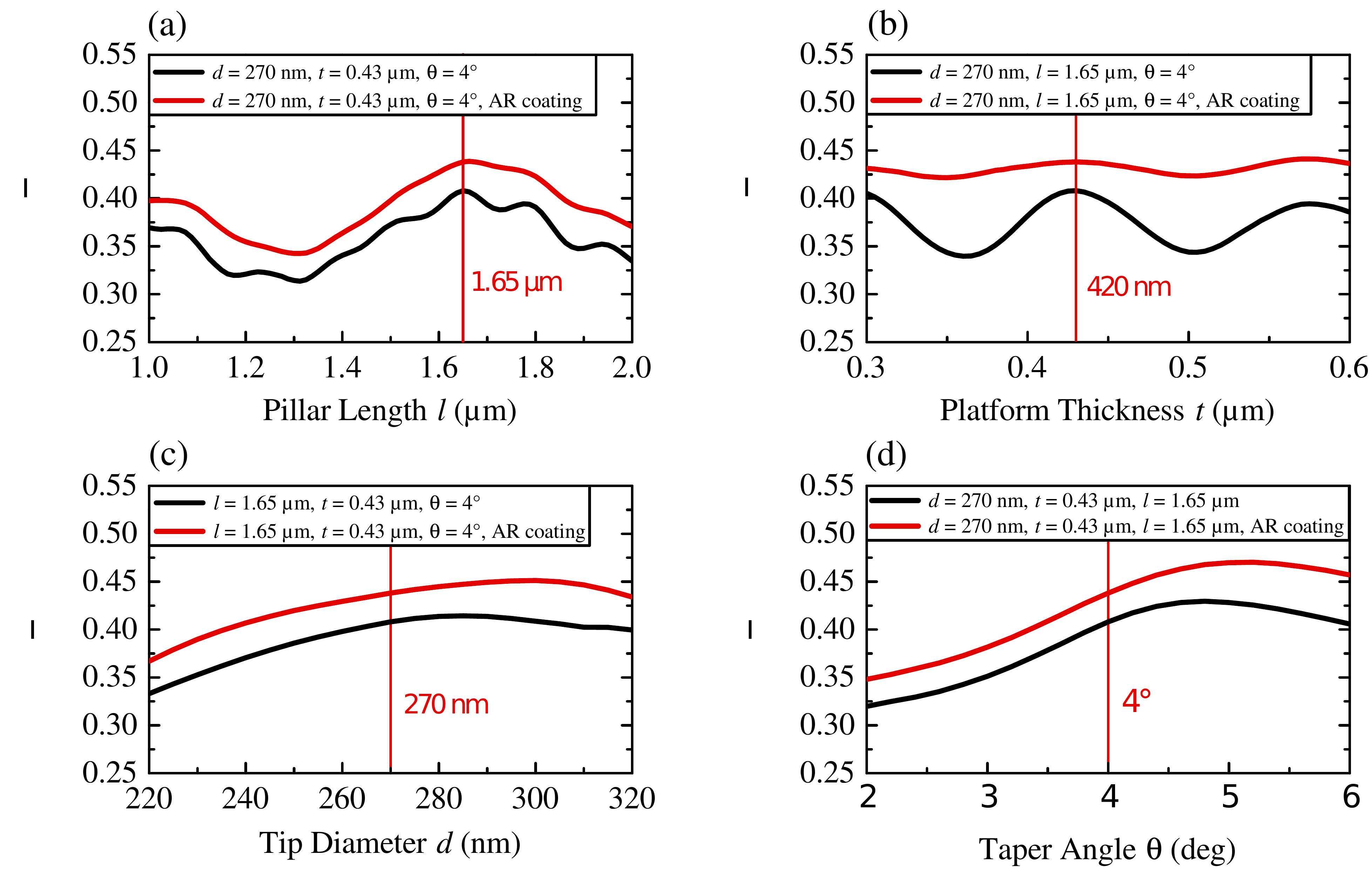}
\caption{We find the optimal device geometry for a pillar with length $l=1.65~\mu$m (a), a platform thickness of $t=0.43~\mu$m (b) and a top diameter of $d=270$ nm (c), together with a taper angle of $\theta = 4^\circ$. Shown are sweeps of a single parameter, while all other parameters remained fixed at their optimal value. The black lines depict the situation without any AR coating on the platform, the red lines are simulated with an AR coating of 122 nm of silica on the platform.}
\label{fig:StepByStep}
\end{figure}

In the following, we discuss the fabrication tolerances for our optimal device geometry. To do so, we sweep again each parameter separately while the other parameters are fixed to their optimal value. Figure \ref{fig:StepByStep} shows these sweeps for the pillar length $l$ (a), platform thickness $t$ (b), top diameter $d$ (c) and taper angle $\theta$ (d).
We also determine the ideal silica AR coating thickness (for the whole PSB) for the optimal device geometry to 122 nm. Applying this coating yields the red lines in figure \ref{fig:StepByStep} and boosts $\overline{\xi}$ to $0.44$ for our optimal device geometry. 
\par
As discernible from figure \ref{fig:StepByStep} (a), $l=1.65~\mu$m maximizes $\overline{\xi}$. With our ICP-RIE recipe, we obtain pillar etching rates of around $88$ nm/min (compare figure \ref{fig:System_Trenches}), sufficient to achieve accuracies in pillar length of around 20 nm. Within 40 nm deviation from $1.65~\mu$m, $\overline{\xi}$ only decreases from $0.41$ to $0.40$, showing that our accuracy is high enough to achieve a high $\overline{\xi}$.

Sweeping $t$ for the optimal device geometry results again in oscillations, with a local maximum of $\overline{\xi} = 0.41$ at $t=0.43~\mu$m, see figure \ref{fig:StepByStep} (b). However, we found in previous work \cite{Challier2018} that our diamond membranes, from which we sculpt the platforms, possess a thickness gradient with around $1~\mu$m thickness variation over $300~\mu$m lateral distance. Although the platform thickness $t$ is well-defined for single devices, it leads to strongly varying $t$ for different devices depending on their position on the membrane.
Deviations from $t=0.43~\mu$m of $0.1~\mu$m decrease $\overline{\xi}$ already to only $0.34$, reducing also the yield of devices maximizing $\overline{\xi}$. Applying an AR coating would certainly overcome this problem, because it nearly eliminates the influence of $t$ on $\overline{\xi}$. However, we emphasize that also the lowest value for $t$ still results in $\overline{\xi}=0.34$ and thus in a higher collectible photon rate than the starting device ($l=2~\mu$m, $d=200$ nm, $\theta=4^\circ$ and $t=0.5~\mu$m) with $\overline{\xi} = 0.23$.

As already discussed, the top diameter $d$ strongly influences $\overline{\xi}$ via the LDOS. 
We note that for $l=1.65~\mu$m and $t=0.43~\mu$m, a minor enhancement of $\overline{\xi}$ is obtained for $d=285$ nm, whereas our initial optimization with $t=500$ nm gave $d=270$ nm.
Deviations of $\pm 20$ nm basically do not influence $\overline{\xi}$ significantly. With an AR coating, the optimal diameter shifts again to $d=300$ nm, slightly increasing $\overline{\xi}$ to $0.45$. Our EBL generates a beam with a minimal diameter of around 5 nm, that can be scanned over the sample with 6 nm step size, yielding an estimated EBL accuracy of around 10 nm. However, over- and underexposure of our mask can lead to variations in the actually achieved top diameter. Thus, we can currently only coarsely estimate the accuracy to be in the order of 50 nm. 

Lastly, our most probable taper angle of $\theta=4^\circ$ is not the optimal value, as can be seen from figure \ref{fig:StepByStep} (d), but still significantly better than a vanishing taper angle. If we could tune the taper angle separately with an accuracy in the order of $1^\circ$, one could certainly optimize $\overline{\xi}$ further. 
Staying at $\theta=4^\circ$, the optimal device geometry found here describes a feasible way to tune our devices towards a nearly twofold increment of the collectible photon rate, significantly enhancing $\eta_\text{DC}$ using our well established nanofabrication process.
\section{Summary \& Outlook} 
In this work, we simulated NV based scanning probe devices and identified the mechanisms influencing the collectible photon rate, which we quantified with the average collection factor $\overline{\xi}$.
First, we investigated the influence of different orientations and lateral  positions of the emitting dipole inside the pillar: The contribution of the z-dipole to $\overline{\xi}$ is negligible, independently of its lateral position, and a laterally-centered x- or y-dipole maximizes $\overline{\xi}$. Moderate offsets up to roughly 80 nm from the lateral center do not significantly lower $\overline{\xi}$, especially for larger pillar diameters ($d>200$ nm).
Notably, the vertical position does not influence $\overline{\xi}$ significantly within a range suitable for sensing and imaging applications ($h = 2-30$ nm).
\par
In the next step, we focused on the geometric parameters of the pillar. 
The main mechanism behind the influence of the pillar geometry on $\overline{\xi}$ are standing waves forming inside the pillar, defined by the pillar length $l$ and the effective mode index $n_\text{eff}$. Depending on the position of their nodes and anti-nodes relative to the dipole, the LDOS and thereby $\overline{\xi}$ enhances or reduces. 
The taper angle and diameter of the pillar are thus probably the most important parameters, because they determine the confinement of the \HE modes and the effective mode index $n_\text{eff}$. For $d < 250$ nm, the confinement improves for increasing $d$ and converges to unity for $d>250$ nm, maximizing the field strength of the \HE modes inside the pillar. Sweeping the length $l$ subsequently tunes an anti-node to the position of the dipole, maximizing the LDOS and $\overline{\xi}$. However, we found these effects, which are strong for a single wavelength, to blur when averaged over the 100 nm broad sideband emission of the NV center, which enhances the tolerances for nanofabrication. If we consider the realistic case of a tapered pillar ($\theta \approx 4^\circ$), $\overline{\xi}$ increases compared to an untapered pillar, rendering tapered pillars favorable.
\par
The platform forms a decoupled, dielectric layer, trough which the NV fluorescence passes to reach the upper half space and the collection optics.  A varying Frensnel transmission leads to an oscillatory behaviour of $\overline{\xi}$ for varying thickness of the platform, whereas an AR coating could reduce this effect and boost $\overline{\xi}$ further. 

Putting all these findings together enables us to propose an optimized device geometry, which maximizes $\overline{\xi}$ for our currently achieved taper angle of $\theta = 4^\circ$: A pillar length of $l=1.65~\mu$m, top diameter of $d=270$ nm and a platform thickness of $t=0.43~\mu$m boosts our device to $\overline{\xi} = 0.41$, thus enhancing the collectible PL rate by a factor of $13$ compared to $\overline{\xi} \approx 0.03$ for bulk diamond and a factor of $1.8$ compared to our non-optimized geometry with $\overline{\xi}=0.23$. With an AR coating on the platform, we could even overcome limits in fabrication accuracy and further increase the performance of our device to $\overline{\xi} = 0.43$.
\par
Even though we already considered special geometry features, including tapered pillars and the formation of trenches, other tip geometries, e.g.\ a spherically or parabolically-shaped tip, could further enhance $\overline{\xi}$. Mask erosion on purpose could also be used to fabricate pillars which possess both a tapered part at the base and a straight part towards the tip. Anyway, we want to finally emphasize that the optimizations done in this work aim at a higher collectible photon rate within the boundaries given by the application, which requires the nanopillar to form a suitable tip for nanoscale sensing. Dropping these boundaries might increase $\overline{\xi}$ further, aiming at highly-efficient single photon sources, which have already been demonstrated using cylindrical pillars \cite{Babinec2010}.
\section{Acknowledgements} 
We gratefully acknowledge funding via a NanoMatFutur grant of the German Ministry of Education and Research (FKZ13N13547) as well as a PostDoc Fellowship by the Daimler and Benz Foundation. We thank Dr.\,Rene Hensel and Susanne Selzer (INM, Saarbrücken) for providing the plasma etching tool and assistance, Thomas Veit for enabling to use the AFM device.
\section{References}
\bibliographystyle{iopart-num}
\bibliography{NV}

\end{document}